\documentclass[aps,reprint,twocolumn,groupedaddress,amsmath,amssymb]{revtex4-1}

\usepackage{graphicx}
\usepackage{bm}
\usepackage{amsmath,hyperref}
\usepackage{amssymb}
\usepackage{amsfonts}
\usepackage{fancyhdr}
\usepackage{slashed}
\usepackage{latexsym,bbm}
\usepackage{amsthm}
\usepackage{float}
\usepackage{amsthm}
\usepackage{tikz}
\usetikzlibrary{matrix,shapes,arrows,positioning,chains}
\usepackage{algorithm,algpseudocode}
\usepackage{enumerate}

\newcommand{\ket}[1]{\left |{#1} \right \rangle}

\usepackage{color}
\definecolor{blue}{rgb}{0,0.2,1}

\definecolor{red}{rgb}{0.9,0,0}

\newcommand{\Ord}[1]{\mathcal{O}\left( #1 \right)}
\newcommand{\tOrd}[1]{\tilde{\mathcal{O}}\left( #1 \right)}

\theoremstyle{plain}

\newtheorem{theorem}{Theorem}

\newtheorem{lemma}{Lemma}

\newtheorem{defn}{Definition}

\def\be{\begin{eqnarray}}
\def\ee{\end{eqnarray}}

\begin{document}

\title{Quantum algorithm for Markov Random Fields structure learning by information theoretic properties}
\author{Liming Zhao}\email{zlm@swjtu.edu.cn}
\thanks{The School of Information Science and Technology, Southwest Jiaotong University, Chengdu 610031, China}
\author{Lin-chun Wan}
\thanks{School of Computer and Information Science, Southwest University, Chongqing 400715, China}
\author{Ming-xing Luo}\email{mxluo@home.swjtu.edu.cn}
\thanks{The School of Information Science and Technology, Southwest Jiaotong University, Chengdu 610031, China}

\date{ }
\begin{abstract}
Probabilistic graphical models play a crucial role in machine learning and have wide applications in various fields.
 One pivotal subset is undirected graphical models, also known as Markov random fields.
In this work, we investigate the structure learning methods of Markov random fields on quantum computers. We propose a quantum algorithm for structure learning of an $r$-wise Markov Random Field with a bounded degree underlying graph, based on a nearly optimal classical greedy algorithm.  The quantum algorithm provides a polynomial speed-up over the classical counterpart in terms of the number of variables. Our work demonstrates the potential merits of quantum computation over classical computation in solving some problems in machine learning. 
\end{abstract}

\maketitle 

\section {Introduction}

 Probabilistic graphical models have been extensively studied since Pearl proposed \cite{pearl1988probabilistic}  and are widely used in probability theory and machine learning \cite{friedman2004inferring,larranaga2011probabilistic,farasat2015probabilistic,larranaga2012review,bacchus2013graphical}. A graphical model can be represented by a graph where each node corresponds to a variable and each edge describes the dependency correlation.  A Markov random field (MRF) is over an undirected graph that describes a set of random variables having Markov properties.  MRFs have a wide range of applications  such as statistical physics, computer vision, machine learning and computational biology \cite{geman1986markov, clifford1990markov,diebel2005application,ma2014mrfalign}. 
Learning an MRF is to recover the underlying graph and the interaction parameters, using a number of samples. Until recently, many algorithms for learning MRFs have been constructed \cite{mckenna2019graphical, klivans2017learning, hamilton2017information}. 
An important problem in MRF learning is structure learning, which is learning the neighbors of each node of the underlying graph.  The structure learning problem has been studied for some  MRFs with special structures such as tree graph \cite{chow1968approximating}, polytrees \cite{dasgupta1999learning}, hypertrees \cite{srebro2003maximum}, and tree mixtures\cite{anandkumar2012learning}.  
In general, 
structure learning of MRF is hard, even for MRFs defined on graphs with a bounded degree \cite{bresler2014structure,klivans2017learning}.
  As the dimensionality of the data to be processed  increases continuously, it gradually exceeds the capabilities of classical computers. Quantum computing provides a way to tackle these problems.

 It has been shown that quantum algorithms solve certain problems much more efficiently than the corresponding classical algorithms \cite{montanaro2016quantum}, such as various problems in machine learning \cite{biamonte2017quantum}.  Quantum computation is also promising to perform much better in learning structures of MRFs than classical computation. In Ref.~\cite{rebentrost2021quantum}  \citeauthor{rebentrost2021quantum} introduced a quantum algorithm for  Ising models which are special MRFs with pairwise interactions and proved that the quantum algorithm is more efficient than the classical counterpart \cite{rebentrost2021quantum}.  In Ref.~\cite{zhao2021quantum}, a quantum algorithm for structure learning of $r$-wise MRFs with  bounded degree graphs has been constructed with run time $\Ord{n^{{r/2+1/2}}}$ while the run time of the classical algorithm is $\Ord{n^r}$ \cite{klivans2017learning}.  
Here $r$-wise means there are at most $r$-nodes contained in each clique, where a clique is a fully-connected subgraph of the underlying graph.

In this work, we consider constructing a quantum algorithm for structure learning of $r$-wise MRFs with bounded degree $d$ based on the classical algorithm in Ref.~\cite{hamilton2017information}. The classical greedy algorithm requires samples which doubly exponential in $d$ while the  algorithm by Sparsitron in Ref.~\cite{klivans2017learning} is exponential in $r$. The advantage of the former one is that it can be used to learn a model only partial observations are allowed or a model where each node is erased independently with a fixed probability $p$.  Therefore, it is worth considering designing an efficient quantum version of it.  We make a specific version (fix the search order) of the classical algorithm  and construct a quantum algorithm accordingly. We also prove that the quantum algorithm provides polynomial speed-up over the classical counterpart, in terms of the dimension of samples. Here only the binary MRFs are considered, as it is easy to generalize to non-binary MRFs. 
 
The paper is arranged as the following. In Sec.~\ref{Preliminary}, we define the notations used  and introduce the  $(\alpha, \beta)$-non-degenerate Markov random fields. In Sec.~\ref{section_classical}, we show the classical algorithm for structure learning of $r$-wise MRFs with bounded degree $d$ underlying graphs. The quantum algorithm and the run time are then presented In Sec.~\ref{section_quantum}. We finally give the summary in Sec.~\ref{section_conclusion}. 

\section{Preliminaries}\label{Preliminary}
 
 \subsection{Notations}
 
 Let $\mathbb R$ denote the set of real numbers,  $\mathbb Z_+$ denote the set of positive integers and  $[N]=\{1,2,\cdots,N\}$. Denote each variable as a capital letter and each configuration of a variable as a lowercase letter. Given a variable $X$ and a number $x$, define an indicator function 
\begin{eqnarray}
\mathbbm{I}_{\{X=x\}} = 
\begin{cases}
1, &  X=x\\
0, &  X \neq x.
\end{cases}
\end{eqnarray}
 All logarithms are base $2$ and denoted by $\log$. If an algorithm costs time  $f(n) = O(n \log^k n)$ for some positive constant $k$,  we denote the run time as $\tOrd{n}$.  Denote $\mathbb E[X]$ as the expectation of variable $X$ and $\widehat{\mathbb E}[X]$ as the empirical expectation of variable $X.$ Denote  the set $[n]$ without $u$ as $[n]\setminus \{u\}$.
 For any quantum state $\ket{ \Psi } $, denote the tensor product of $M$ such states as $\ket{\Psi}^{\otimes M }$.  For states $\ket{a_k}(k\in[M])$, denote the tensor product  $\ket{a_1}\cdots \ket{a_M}$ as $\ket{a_k}^{\otimes_{k=1}^M}.$
 
\subsection{Markov random fields }
 
 A Markov random field is an undirected graphical model.  An undirected graph  can be denoted as $G(V, E)$ where $V$ is the vertex set and $E$ is the edge set.  A clique $C$ of graph $G$ is defined as a subset of vertices that are fully connected.  Each clique is also known as a hyperedge. Let $\mathcal C$ as the set of all cliques.  An MRF is called $r$-wise if each clique (hyperedge) of the underlying graph contains at most $r$ nodes.
Here we consider binary MRFs with $n$ nodes only where each node being a variable $X_j \in \{-1,1\} \left( j\in [n] \right)$.
The MRF is defined by a joint probability distribution $\mathcal D$ on $X \in \{-1,1\}^n$ 
\begin{eqnarray}
p(x) =\frac{1}{Z} \exp\left(\sum_{l=1}^r\sum_{\{i_1<i_2\cdots< i_l\}\in \mathcal C} \theta^{i_1\cdots i_l} \left(x_{i_1},\cdots , x_{i_l} \right)\right),\nonumber\\
\end{eqnarray}
 where tensor $\theta^{i_1\cdots i_l}  : \mathbb R^l \to \mathbb R$ is a function for clique $\left(i_1,\cdots,i_l\right)$, $i_k\in[n]$ for each $k\in[l]$, $Z$ is a normalization factor and known as partition function.
 \begin{figure}[h!]
  \includegraphics[width=0.7\linewidth]{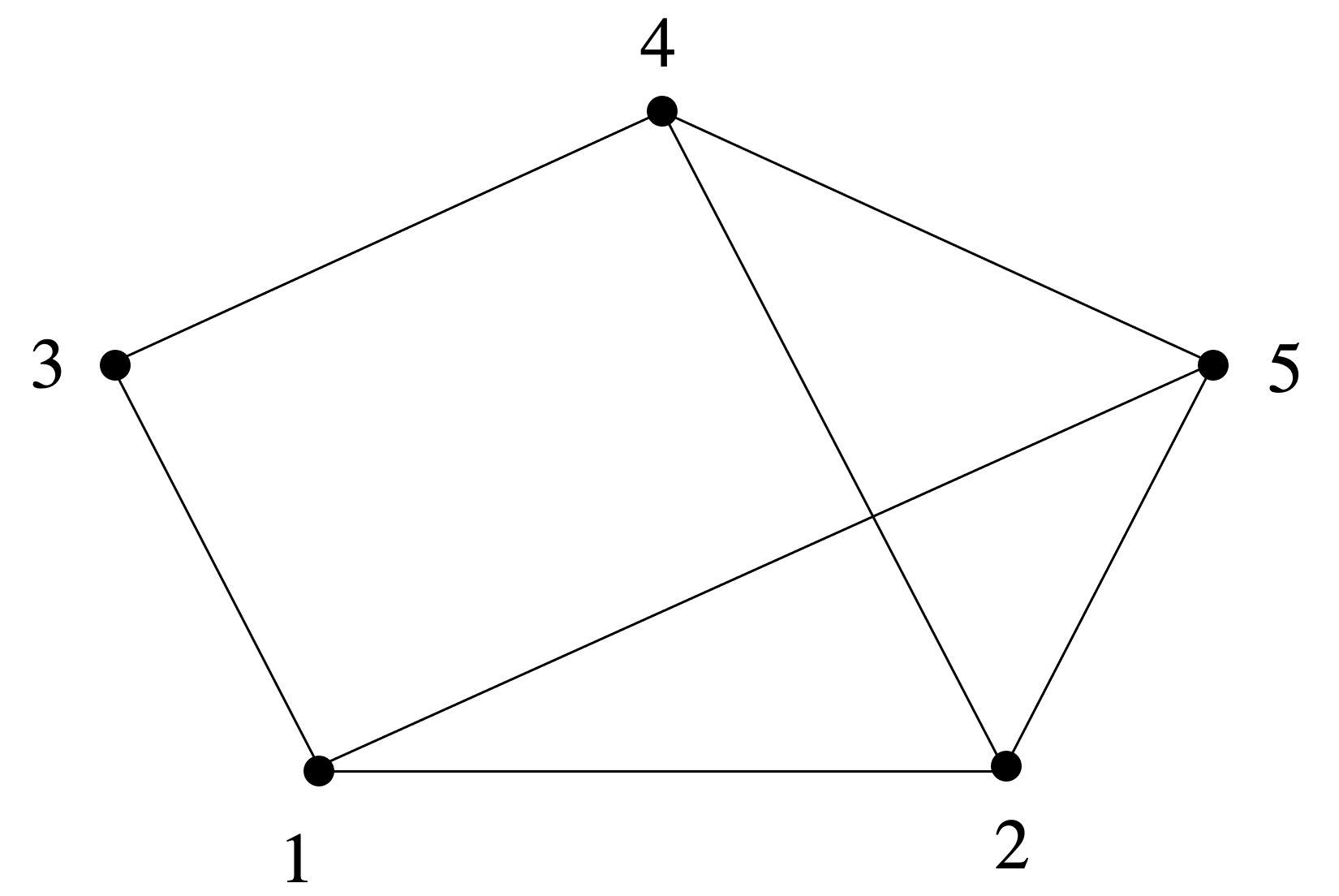}
  \caption{The underlying graph of a $5$-variable MRF}\label{figure_MRF}
\end{figure}

We take an MRF with  a $5$-node underlying graph  shown in Figure  \ref{figure_MRF} as an example. There are two $3$-node cliques $\left(1,2,5\right)$ and $\left( 2,4,5\right)$,  seven $2$-node cliques $\left(1,2\right)$, $\left(1,3\right)$, $\left(1,5\right), $ $\left(2,4\right)$, $\left(2,5\right)$, $\left( 3,4\right)$, $\left( 4,5\right)$, and five $1$-node cliques. Let all the  $3$-node cliques constructing set $\mathcal C_3$ while all $2$-node cliques constructing set $\mathcal C_2$. All cliques form set $\mathcal C$.  The joint probability distribution of the MRF with the underlying graph shown in Figure \ref{figure_MRF} can be written as
 \begin{eqnarray}
 p(x) &=&\frac{1}{Z} \exp\left(\sum_{l=1}^3\sum_{\substack{\{i_1<i_2\cdots< i_l\}\in \mathcal C}} \theta^{i_1\cdots i_l} \left(x_{i_1},\cdots , x_{i_l} \right)\right),\nonumber\\
 &=&\frac{1}{Z} \exp \left(\sum_{i_1=1}^5  \theta^{i_1} \left(x_{i_1} \right) + \sum_{\{i_1<i_2\}\in \mathcal C_2} \theta^{i_1i_2} \left(x_{i_1}, x_{i_2} \right) \right. \nonumber\\
 &  & \left. + \sum_{\{i_1<i_2<i_3\}\in \mathcal C_3} \theta^{i_1i_2i_3} \left(x_{i_1}, x_{i_2}, x_{i_3} \right)\right).
 \end{eqnarray}
Here the MRF is a $3$-wise MRF because each clique contains at most three nodes. Every clique with three nodes is called a maximal clique of the MRF in Figure \ref{figure_MRF}.  For any MRF, the definition of the maximal clique (hyperedge) of the underlying graph is given in the following.

 \begin{defn}[Maximal clique]
  A clique is a maximal clique if  it is not contained in any other cliques with more nodes. For a maximal clique $J$, there is no clique $I\subset [n]$ such that $J\subset I$. A maximal clique is also called a maximal hyperedge. 
\end{defn} 

For example, in Figure \ref{figure_MRF}, other than the $3$-node cliques $\left(1,2,5\right)$, $\left(2,4,5\right)$, the $2$-node cliques $\left(1,3\right)$,  $\left(3,4\right)$ are maximal cliques (hyperedges) as well.  All the cliques in $\mathcal C_2 \setminus \{ \left( 1,3\right),  \left(3,4\right)\}$ are not maximal cliques as each clique in this set is contained in either $\left(1,2,5\right)$ or $\left(2,4,5\right).$

Let $H$ be the set of hyperedges of the underlying graph $G$ of an MRF. In order to ensure the existence and absence of each maximal hyperedge is information-theoretically determined, assume an MRF satisfies the following non-degeneracy condition. 
 \begin{defn}[Definition 2.2 in Ref.~\cite{hamilton2017information}]\label{defn_alpha_beta_nondegenerate}
 A Markov random field is $(\alpha, \beta)$-non-degenerate if
 \begin{enumerate}
 \item Every edge $\left(i,j\right)\in E $ in the underlying graph is contained in some hyperedge $h\in H$  where the corresponding tensor is non-zero.
 \item  Every maximal hyperedge $h\in H$ has at least one entry lower bounded by $\alpha$ in absolute value.
 \item  Every entry of  $ \theta ^{i_1i_2\dots i_l}$ is upper bounded by a constant $\beta$ in absolute value.
 \end{enumerate}
 \end{defn}
 
We  illustrate  Definition \ref{defn_alpha_beta_nondegenerate} by taking the Figure \ref{figure_MRF} as an example. Assume Figure \ref{figure_MRF} is the underlying graph of an  $(\alpha, \beta)$-non-degenerate MRF, then it should satisfy the above three constraints. We first explain the first constraint by an edge in a $2$-node hyperedge $(1,3)$ and an edge in a $3$-node hyperedge $(1,2,5)$.  Edge $(1,3) $ is only contained in hyperedge $(1,3)$, then $\theta^{13}\neq 0$.  Edge $(1,2)$ is in both hyperedge $(1,2)$ and $(1,2,5)$, then according to the first constraint, either $\theta^{12}$ or  $\theta^{125}$ is non-zero.  If the maximal hyperedge $\left( 1,2,5\right)$ satisfies the second constraint, that means $\max\{ \vert \theta^{125} \vert, \vert \theta^{12}\vert, \vert \theta^{15}\vert,\vert\theta^{25} \vert,\vert\theta^{1} \vert, \vert\theta^{2} \vert, \vert\theta^{5} \vert  \}\geq \alpha.$ If the MRF satisfies the third constraint, it indicates that all tensor $\vert  \theta ^{i_1i_2\dots i_l} \vert \leq \beta.$

 We now give the definition of the degree of the underlying graph of an MRF. Denote $N(u)$ as the set of neighbours of vertex $u$ of the underlying graph $G$. The number of elements in $N(u)$ is defined as the degree of vertex $u$ which we denote as $d_u$. 
\begin{defn}
  We say the underlying graph $G$ of an MRF is $d$-degree bounded if $d=\max_{u\in[n]} \{d_u\}$.
\end{defn} 
For example, in Figure \ref{figure_MRF}, the neighbours of node $1$ is $ N(1)=\{2,3,5\}$,  the degree of node $1$ is  $d_1=3$ and   the degree of the whole graph is $d=3$ as $\max_{i\in[5]} \{d_i\}=3.$
 
For an MRF, the conditional distribution probability of a node $u$ only depends on the state of neighbours of $u$,  i.e.\   it satisfies the Markov property which can be expressed as
 \begin{eqnarray}
  p\left(x_u\vert x_{[n]\setminus \{u\}}\right) =  p\left(x_u\vert x_{N(u)} \right),
 \end{eqnarray}  
 where $p\left(x_u \vert x_I\right)$ is short for  $p\left(X_u=x_u \vert X_I=x_I\right)$ for $I\subset [n]\setminus \{u\}$.

\subsection{Quantum subroutines}

The main subroutine used in this paper is the quantum maximum finding algorithm  \cite{durr1996quantum}, by which we can find the maximum number in a given data set. It is quadratically faster than the corresponding classical algorithms. 
\begin{lemma}[Quantum maximum finding \cite{durr1996quantum}]\label{quantum_max_finding}
Given quantum access to a data set $X$ of size $N$, we can find the maximum element  with success probability $1-
\eta $ with $\Ord{\sqrt{N}\log(1/\eta)}$ queries.
\end{lemma}

 \section{Learning structure of  MRFs via mutual information}\label{section_classical}
 In Ref~\cite{hamilton2017information}, the authors proposed a scheme for learning structures of $r$-wise, $(\alpha,\beta)$-non-degenerate MRFs with degree $d$ underlying graph  based on some information properties. The scheme is to learn the whole graph via learning the neighbours of each node, for a given number of samples from the distribution $\mathcal D$ of an MRF.  
 
 For any node $u$, and a subset $S$ which is a subset of the neighbours of $u$ but does not contain all the neighbors of $u$, there is a lower bound of the conditional mutual information (conditioned on $X_S$) between node $u$ and a subset $I$ which contains at least one of the neighbors of $u$ \cite{hamilton2017information}, see Lemma \ref{lemma_mutual_information}.  Based on the lower bound, a greedy algorithm (named as algorithm \textsc{MfrNbhd}) to learn the neighbours of the node $u$ was proposed in  Ref~\cite{hamilton2017information}. In this section, we introduce  the algorithm \textsc{MfrNbhd} with a specific search order.

 \subsection{An information quantity related to mutual information and the lower bound}
 
 Given a number of samples, the neighbours of a fixed node $u$ can be found via calculating an information quantity related to the mutual information and using a threshold constant (the lower bound mentioned above).  The definition of the related information quantity is given in the following. 
\begin{defn}
For a node $u\in[n]$ and subsets $S\subset [n]\setminus \{u\}$ and $I\subset [n]\setminus \{S\cup \{u\}\}$, define a certain information theoretical quantity, 
  \begin{eqnarray}
 {v}_{u,I \vert S}  
:= \mathbb E_{\{\substack{x_u,\\x_I}\}} \mathbb{E}_{x_S}\left[\vert {p}\left(x_u, x_I\vert x_S\right) -{p}(x_u) {p}\left( x_I\vert x_S\right)\vert \right].\nonumber
 \end{eqnarray}
 \end{defn}
 For $M$ samples of an MRF, the empirical version of ${v}_{u,I \vert S}$ is given by
  \begin{eqnarray}
\hat{v}_{u,I \vert S} := 
 \mathbb E_{\{\substack{x_u,\\x_I}\}} \hat{\mathbb E}_{x_S}\left[\vert\widehat{p}\left(x_u, x_I\vert x_S\right) -\widehat{p}(x_u) \widehat{p}\left( x_I\vert x_S\right)\vert \right],\nonumber\\ 
 \label{equ_mrf_information}
 \end{eqnarray}
 where $$\widehat{p}(x_{I'}) =\frac{1}{M}\textstyle\sum_{m=1}^M \mathbbm{I}_{\{x_{I'}^{(m)}=x_{I'}\}}$$
 is the empirical probability of configuration $x_{I'}$ of subset $I'$ in the $M$ samples.

 We now give the lower bound of the information-related quantity defined above. The lower bound is also called the threshold. For any node $u\in[n]$ of an $r$-wise MRF and its neighbours $N(u)$, let  
 \begin{eqnarray}
 \gamma & := &\sup_u  \sum_{l=1}^r\sum_{i_2<\cdots<i_l}  \vert \theta^{ui_2\cdots i_l} \vert_{\max} \leq \beta\sum_{l=1}^r \tbinom{d}{l-1}, \nonumber\\
 \delta &:=& \frac{1}{2}\exp(-2\gamma)  
 \end{eqnarray}
 where $\vert \theta^{ui_2\cdots i_l} \vert_{\max} $ is the maximum entry of the tensor. The threshold constant is then defined as  \begin{eqnarray}
 \tau:=\frac{2\alpha^2\delta^{r-1}}{r^{2r}2^{r+1} \binom{d}{r-1}\gamma e^{2\gamma}}.\label{equation_tau}
\end{eqnarray}

 \begin{lemma}[Theorem C.6 in Ref.~\cite{hamilton2017information}]\label{lemma_mutual_information}
For a  vertex $u$ of an $(\alpha,\beta)$-nondegenerate MRF, and a subset of the vertices $S\subsetneq N(u)$.  Then taking a set $I$ uniformly at random from the subsets of $N(u)\setminus S$ of size $\kappa=\min(r-1, |N(u) \setminus S|)$, we have
\begin{eqnarray}
 \mathbb E_I\left[\sqrt{\frac{1}{2}  I\left(X_u;X_I\vert I_S\right)}\right] \geq \mathbb E_I\left[v_{u,I|S}\right] \geq 2\tau.
\end{eqnarray}
\end{lemma}
It has been shown that $v_{u,I|S}$ and $\hat{v}_{u,I|S}$ are very close when given enough samples, and a subset $I$ which contains vertices in $N(u)\setminus S$ can be found by using Lemma \ref{lemma_mutual_information}, if $S$ do not contain all nodes of the neighbours of node $u$.  Specifically,  the entire neighbours of any node $u$ can be found via finding all subsets $I$ with size at most $r-1$ such that the empirical information quantity satisfies
\begin{eqnarray}
\hat{v}_{u,I\vert S} > \tau.\label{eq_threshold_v_by_tau}
\end{eqnarray}

\subsection{Classical algorithm for MRF structure learning}

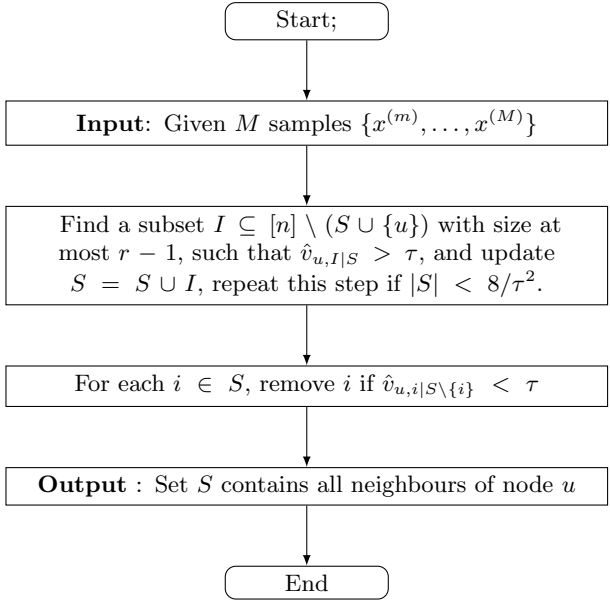
\begin{figure}[h!]\label{Mrf_Nbhd}
\begin{center}
\tikzset{
decision/.style={
    diamond,
    draw,
    text width=4em,
    text badly centered,
    inner sep=0pt
},
block1/.style={
    rectangle,
    draw,
    text width=6em,
    rounded corners
},
block2/.style={
    rectangle,
    draw,
    text width=0.9\linewidth,
},
block3/.style={
    rectangle,
    draw,
    text width=0.9\linewidth,
    text centered,
},
block/.style={
    rectangle,
    draw,
    text width=0.9\linewidth,
    text centered,
},
cloud/.style={
    draw,
    ellipse,
    minimum height=2em
},
descr/.style={
    fill=white,
    inner sep=2.5pt
},
connector/.style={
    -latex,
    font=\scriptsize
},
rectangle connector/.style={
    connector,
    to path={(\tikztostart) -- ++(#1,0pt) \tikztonodes |- (\tikztotarget) },
    pos=0.5
},
rectangle connector/.default=-2cm,
straight connector/.style={
    connector,
    to path=--(\tikztotarget) \tikztonodes
}
}

\begin{tikzpicture}
\matrix (m)[matrix of nodes, column  sep=2cm,row  sep=8mm, align=center, nodes={rectangle,draw, anchor=center} ]{
     |[block1]| {Start};          &  \\
    |[block2]| {\textbf{Input}: Given $M$ samples $\{ x^{(m)},\dots, x^{(M)}\}$ 
    }               &  \\
    |[block3]| { 
    Find  a subset  $I\subseteq [n]\setminus (S\cup \{u\})$ with size at most $r-1$, such that  $\hat{v}_{u,I\vert S}>\tau$, and update $S=S\cup I$, repeat this step if $\vert S \vert <8/\tau^2.$ 
        }         &                                            \\                                     
  |[block3]| {For each $i\in S$,  remove $i$ if $\hat{v}_{u,i \vert S\setminus \{i\} } < \tau$}    &                                             \\                                     
  |[block]| {\textbf{Output} : Set $S$ contains all neighbours of node $u$}                         &                                             \\
  |[block1]| {End}          &  \\
};
\path [>=latex,->] (m-1-1) edge (m-2-1);
\path [>=latex,->] (m-2-1) edge (m-3-1);
\path [>=latex,->] (m-3-1) edge (m-4-1);
\path [>=latex,->] (m-4-1) edge (m-5-1);
\path [>=latex,->] (m-5-1) edge (m-6-1);
\end{tikzpicture}
 \caption{Algorithm \textsc{MrfNbhd} \cite{hamilton2017information} }\label{figure_algorithm_framework}
\end{center}
\end{figure}

Given a number of samples from an $r$-wise, $(\alpha,\beta)$-non-degenerate MRF with degree $d$ underlying graph, fix a node $u\in [n]$,  there is an algorithm which recover the neighbours of $u$ \cite{hamilton2017information}. The framework of the algorithm is shown in Figure \ref{figure_algorithm_framework}.

The key step of the algorithm in Ref.~\cite{hamilton2017information} is to find a subset $I$ with size at most $r-1$ which satisfies the constrain in Eq.~(\ref{eq_threshold_v_by_tau}). 
There is no constraint on the search order. Here we use a specific search order for constructing a quantum algorithm more straightforwardly. The quantum version will be introduced in Section \ref{section_quantum}.   We divide the search step into $r-1$ search round at most and fix the size of  potential subsets to be searched at each search round. More details, for an $r$-wise MRF, fix a node $u$ and a set $S\subset [n]\setminus \{u\}$,  there are  $K=\sum_{l=1}^{r-1}  \tbinom{n-1-\vert S \vert }{l} $ different subset $I\subseteq [n]\setminus (S\cup \{u\})$. Here we divide the $K$ cliques into $r-1$ classes by the sizes of the cliques and the $l$-th class contains $K_l=  \tbinom{n-1-\vert S \vert }{l}$ cliques. Define these sets in the following. 
\begin{defn}\label{defn_subset_F}
Fix a node $u$ and a set $S\subset [n]\setminus \{u\}$, define
\begin{itemize}
\item $ F_{u,S}=\{I~\vert ~I\subseteq [n]\setminus (\{u\}\cup S), \vert I \vert \leq r-1\}.$
\item $ F^{l}_{u,S}=\{I~\vert ~I\in F_{u,S}, \vert I \vert = l\},$ for $l\in [r-1].$
\end{itemize}
\end{defn}

The classical structure learning algorithm with a fixed search order is shown in Algorithm \ref{MRF_information}.   The correctness of the algorithm and the sample complexity has been proved in Ref.~\cite{hamilton2017information}, where the number of samples is given by
 \begin{eqnarray}
M &\geq &\frac{60*2^{2L}}{\tau^2\delta^{2L}}(\log\frac{1}{w}+\log(L+r)+(L+r)\log(2n)+1),\nonumber
\end{eqnarray}
 where $L=8/\tau^2.$ For the sake of convenience, we use sample complexity based on the above inequality  
\begin{eqnarray}
M &\in & \tOrd{ \frac{2^{2L}}{\tau^2\delta^{2L}}\left(\log\frac{1}{w}+(L+r)\log n\right)}\nonumber\\
&= & \tOrd{ \frac{2^{2L}}{\tau^2\delta^{2L}}\left(\log\frac{1}{w}+\log n\right)}
\end{eqnarray}
where the last equation is obtained by the fact that $2^r$ is a factor of $1/\tau$.

The sample complexity and the run time of Algorithm \ref{MRF_information} are given in the following theorem. 
\begin{theorem}[Theorem 5.7 in Ref.~\cite{hamilton2017information}]
Fix $w > 0$. Suppose we are given $M$ samples from an $(\alpha,\beta)$-non-degenerate Markov random field over $n$ variables with $r$-order interactions where the underlying graph has degree $d$. Suppose that
 $$M\in \tOrd{\frac{2^{2L}}{\tau^2\delta^{2L}}\left(\log \frac{1}{w}+\log n\right)},$$
 where $L=\frac{8}{\tau^2}$,  $\tau$ is defined in Eq.~(\ref{equation_tau}), and $L$ is the upper bound of the size of the superset of sets $S\subseteq N(u)$. 
Then with probability at least $1-w$, run Algorithm \ref{MRF_information} on node $u$ recovers the correct neighborhood of $u$, and thus recovers the underlying graph $G$ when run the algorithm on each node. Furthermore, recovering the underlying graph takes $\tOrd{Mn^r}$ time.
 \end{theorem}
 
 \begin{figure}[htbp]
\begin{algorithm}[H]
  \caption{    \textsc{AmcvMfrNbhd}} \label{MRF_information}
  \begin{algorithmic}[1]
  \Require{Training set $\{X^{(m)} \in \{-1,1\}^n\} (m\in M),$ a fixed node $u$, and $L, \tau>0$}
    \State  Set $S:=\emptyset$.
   \While{$|S| \leq L$ }
         \State{$l=r-1$ }
         \While{$l >1$ }
    \If {find a set of vertices  $I \in F_{u,S}^l$ defined in Definition \ref{defn_subset_F}, such that $\hat{v}_{u,I\vert S} > \tau$, }\label{step_classical_search}
       \State  set $S:=S\cup I$ 
       \State go to step $2$
        \EndIf  
        \State $l=l-1$   
    \EndWhile
    \EndWhile
    \State For each $i \in S$, if   $\hat{v}_{u,i \vert S\setminus i }< \tau$,   then remove $i$ from $S$.
      \Ensure{ Set $S$  as our estimate of the neighborhood of $u$.}
  \end{algorithmic}
\end{algorithm}
\end{figure}

Here we give a brief analysis of the time complexity of Algorithm \ref{MRF_information}. For any node
$u$, the overload step is Step \ref{step_classical_search}, obtaining  $\hat{v}_{u,I\vert S}$ for  $ \tbinom{n}{l} \in \Ord{n^{r-1}} (l\leq r-1)$  different subsets $I$ of the vertices in $[n] \setminus S$ of size $l$ requires $\Ord{M n^{r-1}}$ time. 
Thus it takes $\tOrd{rMLn^{r-1}}$ time in at most $r-1$ of the inner while loop and total $L$ iterations for the outer while loop. The run time is  then $\tOrd{rMLn^r}=\tOrd{Mn^r}$ time in total for find the neighbours of all $n$ nodes, here factor $r$ and $L$ is removed because $2^r$ and $2^{2L}$ are factors of $M$.

\section{Quantum algorithm for MRF recovering via mutual information}\label{section_quantum}

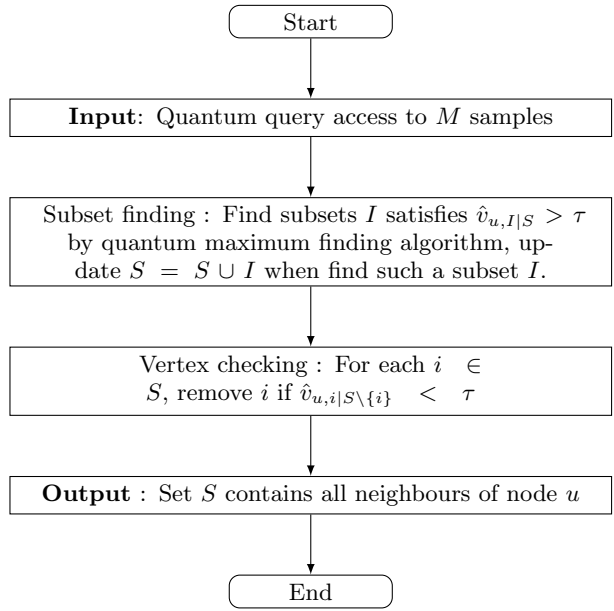
\begin{figure}[h!]
\begin{center}
\tikzset{
decision/.style={
    diamond,
    draw,
    text width=4em,
    text badly centered,
    inner sep=0pt
},
block1/.style={
    rectangle,
    draw,
    text width=6em,
    rounded corners
},
block2/.style={
    rectangle,
    draw,
    text width=0.9\linewidth,
},
block3/.style={
    rectangle,
    draw,
    text width=0.9\linewidth,
    text centered,
},
block/.style={
    rectangle,
    draw,
    text width=0.9\linewidth,
    text centered,
},
cloud/.style={
    draw,
    ellipse,
    minimum height=2em
},
descr/.style={
    fill=white,
    inner sep=2.5pt
},
connector/.style={
    -latex,
    font=\scriptsize
},
rectangle connector/.style={
    connector,
    to path={(\tikztostart) -- ++(#1,0pt) \tikztonodes |- (\tikztotarget) },
    pos=0.5
},
rectangle connector/.default=-2cm,
straight connector/.style={
    connector,
    to path=--(\tikztotarget) \tikztonodes
}
}

\begin{tikzpicture}
\matrix (m)[matrix of nodes, column  sep=2cm,row  sep=8mm, align=center, nodes={rectangle,draw, anchor=center} ]{
     |[block1]| {Start}         &  \\
    |[block2]| {\textbf{Input}: Quantum query access to $M$ samples 
    }               &  \\
    |[block3]| { Subset finding : Find subsets $I$ satisfies $\hat{v}_{u,I\vert S}>\tau$ by quantum maximum finding algorithm, update $S=S\cup I$ when find such a subset $I$. 

        }         &                                            \\                                   
  |[block3]| {Vertex checking : For each $i\in S$,  remove $i$ if $\hat{v}_{u,i \vert S\setminus \{i\} } < \tau$}    &                                             \\                                     
  |[block]| {\textbf{Output} : Set $S$ contains all neighbours of node $u$}                         &                                             \\
  |[block1]| {End}          &  \\
};
\path [>=latex,->] (m-1-1) edge (m-2-1);
\path [>=latex,->] (m-2-1) edge (m-3-1);
\path [>=latex,->] (m-3-1) edge (m-4-1);
\path [>=latex,->] (m-4-1) edge (m-5-1);
\path [>=latex,->] (m-5-1) edge (m-6-1);
\end{tikzpicture}
 \caption{The framework of quantum structure learning algorithm}\label{figure_quantum_algorithm_framework}
\end{center}
\end{figure}

We propose a quantum version of Algorithm \ref{MRF_information} to learn the neighbours of a node $u$ of an $r$-wise, $(\alpha,\beta)$-non-degenerate MRF with degree $d$ underlying graph, and show that it provides a polynomial speed-up over the classical algorithm, in terms of the number of variables. As the classical algorithm,  only the binary MRFs are considered here since the algorithm can be generated to non-binary MRFs straightforwardly.  
The main idea is that assume we are given quantum query access to $M$ samples of such an MRF, we construct a unitary operator $U_v$ to  encode the empirical information quantity in Eq.~(\ref{equ_mrf_information}) into a quantum state which is a superposition of all potential subset $I$ of size $l\in [r-1]$. 
Then use the quantum maximum finding algorithm, which requires performing the  unitary operator $U_v$ a number of times, to find one subset such that the empirical information quantity $>\tau.$  The primary logic is the same as the classical algorithm \ref{MRF_information}. 
The framework of the algorithm is shown in Figure \ref{figure_quantum_algorithm_framework}.

\subsection{Quantum data input} 

In this subsection, we first give the quantum input assumption.  Then for each $l\in [r-1]$, define a string set $\mathcal H_l$ such that there is a unique map from the indexes of a subset $I$ of the nodes with size $l$ to a string in set $\mathcal H_l$. After that, we show  there is a unitary operator  indicating whether a given arbitrary string with size $l$ is in the string set $\mathcal H_l$.

 Assume we are given quantum query access to the samples in the following. 
\begin{defn}[Data input]\label{MRF_input}
Let $x^{(1)},...,x^{(M)}\in\{ 1,-1\}^n$  be $M$ samples from a $n$-variable MRF.  For each $m\in [M]$, assume we are given  quantum query access via the operation
\begin{eqnarray}
Q^{(m)}:~~~~\ket{j}\ket{0}\rightarrow \ket{j}\ket{x_j^{(m)}},
\end{eqnarray}
with time $\Ord{\log n}$, where $j\in[n]$, $m\in [M]$, and  $x_j^{(m)} $ is the value of variable $j$ in the $m$-th sample $x^{(m)}$.
\end{defn}
Notice that the query is a unitary operation, then we have 
\begin{eqnarray}
\ket{j}\ket{x_j^{(m)}} \xrightarrow{Q^{(m)}} \ket{j}\ket{0}.
\end{eqnarray} 

\subsection{Quantum subset representation}

We now consider encoding  subsets $I\subset [n]$ with size $l$ into a quantum state. We map a subset to a string first, then represent the string by a quantum state. As there is no order between the elements in a set,  a mapping from a set to a string is not unique.  We use a string where the elements are placed in increased order to uniquely represent the subset $I$. Define a string set $\mathcal H_l$ in the following.  
\begin{defn}\label{definition_string_set_H_l}
Fix a node $u\in [n]$,  for $l\in[r-1]$,    define a string set $\mathcal H_l$, such that for each string $\mathcal S_j =(j_1,j_2,\cdots,j_l)\in \mathcal H_l$,  it satisfies 
  $$j_1<j_2<\cdots<j_l, $$
where $j_i\in[n]\setminus \{u\}, $ and $i\in [l]$.
\end{defn}
We see that the map from strings in set $\mathcal H_l$ to subsets $I$ with size $l$ is a bijection map.   For any string $\mathcal S_j=(j_1,j_2,\cdots,j_l)$ represented by a quantum state, there is a unitary which indicates whether it is in the set $\mathcal H_l$. The run time is given in the following Lemma. 
\begin{lemma}\label{lemma_string_set_H_l}
Given a string $\mathcal S_j=(j_1,j_2,\cdots,j_l)$, where $j_i\in[n]\setminus \{u\}, \left(i=1,2\cdots,l\right)$, there is a unitary $U_{\mathcal H_l}$ which performs 
\begin{eqnarray}
\ket{0} \ket{\mathcal{S}_j} \to 
\begin{cases}
 \ket{1} \ket{\mathcal{S}_j }, & ~  \mathcal{S}_j \in \mathcal H_l\\
 \ket{0} \ket{\mathcal{S}_j }, & ~ \mathcal{S}_j \notin \mathcal H_l,
\end{cases}
\end{eqnarray}
in time $\Ord{l\log n}$. 
\end{lemma}
\begin{proof}
According to Definition \ref{definition_string_set_H_l},  in order to ensure $j_1<j_2<\cdots<j_l$, the unitary $U_{\mathcal H_l}$ requires $l-1$ comparisons, and each comparison involving $\lceil \log n\rceil$ qubits as each $j_i$ takes $\lceil \log n\rceil$ qubits to represent.  The run time is then $\Ord{l\log n}$. 
\end{proof}

\subsection{Quantum state preparation for the information quantity $\hat{v}_{u,I \vert S}$}

Given quantum access to $M$ samples of an $r$-wise $(\alpha, \beta)$-non-degenerate MRF with degree $d$ underlying graph,  for a subset $I$ in  $ F^{l}_{u,S}$ defined in Definition \ref{defn_subset_F},  we show the empirical information quantity defined in Eq.~(\ref{equ_mrf_information}) can be encoded in a quantum state, which is a superposition of all potential subsets of a fix size $l\in[r-1].$ 

We first expand the empirical information quantity $\hat{v}_{u,I \vert S}$ in  Eq.~(\ref{equ_mrf_information})  as the following 
\begin{eqnarray}
& & \hat{v}_{u,I \vert S}   \nonumber\\
&=&\mathbb E_{\{\substack{x_u,\\x_I}\}} \hat{\mathbb  E}_{x_S} \left[\vert\widehat{p}\left(x_u, x_I\vert x_S\right) -\widehat{p}\left(x_u\vert x_S\right) \widehat{p}\left( x_I\vert x_S\right)\vert \right]\nonumber\\
&=& \textstyle \sum \widehat{p}(x_u) \widehat{p}(x_I)  
\vert  \widehat{p}\left(x_u, x_I,x_S\right) - \widehat{p}(x_u, x_S) \widehat{p}( x_I\vert x_S) \mid \nonumber\\
&=& \textstyle\sum \widehat{p}(x_u) \widehat{p}(x_I) \vert  \widehat{p}\left(x_u, x_I,x_S\right) - \frac{\widehat{p}(x_u, x_S) \widehat{p}( x_I, x_S)}{\widehat{p}(x_S)} \mid, \nonumber\\
\label{eq_expand_version_v}
 \end{eqnarray}
 where the summations are over all configurations of $x_u, x_S, x_I$ in the given samples, and $x_u\in\{-1,1\}, x_I\in\{-1,1\}^{\vert I \vert}$, $x_S\in\{-1,1\}^{\vert S \vert} $ and the empirical probability
 \begin{eqnarray}
 \widehat{p}(x_u) =\frac{1}{M}\textstyle\sum_{m=1}^M \mathbb{I}_{\{X_u^{(m)}=x_u\}.} 
  \end{eqnarray}

We now turn to construct a quantum state to represent  $\hat{v}_{u,I \vert S}$ which is a superposition of all potential subsets $l\in F_{u,S}^l$ in Definition \ref{defn_subset_F}. 
 As $u$ and $S$ are fixed, in the expending of $ \hat{v}_{u,I\vert S}$, the empirical probabilities $\widehat{p}\left(X_u=1\right)$, $\widehat{p}\left(X_u=-1\right)$, $\widehat{p}(x_s)$ and $\widehat{p}\left(x_u,x_s\right)$ can be obtained classically.  Then the main problem is to calculate the terms $\widehat{p}(x_I),$  $\widehat{p}\left(x_I,x_s\right)$,  and $\widehat{p}(x_u,x_I,x_s).$
 
Firstly, we consider preparing the empirical probability $\widehat{p}(x_I)$ of each configuration of any subset $I\in F_{u,S}^l$ in a quantum state. For any subset $I$ with size $l$, there are at most $2^l$ different potential configurations which we denote as  $\{a_1,a_2,\cdots,a_{2^l}\}$. 
The empirical probability of each configuration in $\{a_1,a_2,\cdots,a_{2^l}\}$ can be prepared in a quantum state. The procedure and run time are given in the following.
\begin{lemma}\label{lemma_probability_preparation}
Given quantum access to $M$ samples of an MRF as Definition \ref{MRF_input}, fix a node $u\in [n]$, and $l\in[r-1]$,  a string $\mathcal S_j=\left(j_1,j_2,\cdots j_l\right)\in \mathcal H_l$ corresponding to a subset $I_j \subseteq [n] \setminus\{u\}$ with size $l$, for each potential configuration of $x_{I_j}\in \{a_1,\cdots, a_{2^l}\}$ where $a_k \in \{-1,1\}^l$ for $ (k\in [2^l])$, there is a unitary $U_I$ which performs
\begin{eqnarray}
 \ket{\mathcal{S}_j } \left( \ket{\bar{0}} \ket{\bar{0}}\right)^{\otimes {2^l} }  \to  \ket{\mathcal{S}_j } \left( \ket{a_k} \ket{\widehat{p}\left(x_{I_j}=a_k\right)}\right) ^{\otimes_{k=1}^{2^l} }~~  \label{eq_quantum_state_probability}
\end{eqnarray}
in time $\tOrd{2^{r-1} M+M\log n}$.
\end{lemma}

\begin{proof}

We first show the method to prepare state in Eq.~(\ref{eq_quantum_state_probability}). The procedure is as follows.     
\begin{itemize}
\item[1).]  Prepare an initial state 
\begin{eqnarray}
\ket{\mathcal S_j}\ket{\bar{0}}^{\otimes M}\left(\ket{\bar{0}}\ket{\bar{0}}\right)^{\otimes {2^l}},
\end{eqnarray}
where $\ket{\bar{0}}^{\otimes M}$ is used to store the values of the $M$ samples, $\left(\ket{\bar{0}}\ket{\bar{0}}\right)^{\otimes {2^l}}$ is used to store the potential $2^l$ configurations and the corresponding empirical probability in the $M$ samples. 
\item[2).]  Query access to $M$ samples as Definition \ref{MRF_input} on the first two registers, it yields 
\begin{eqnarray}
\ket{\mathcal{S}_j }\ket{x_{I_j}^{(m)}}^{\otimes_{m=1}^M} \left(\ket{\bar{0}}\ket{\bar{0}}\right)^{\otimes {2^l}} 
\end{eqnarray}
where $I_j$ is the corresponding subset of string $\mathcal S_j$, and  $\ket{x_{I_j}^{(m)}}=\ket{x_{j_1}^{(m)}} \ket{x_{j_2}^{(m)}} \cdots \ket{x_{j_l}^{(m)}}$ stores the value of the configuration of $I_j$ in the $m$-th sample. 
\item[3).] For each $k\in[2^l]$, represent  configuration $a_k$ by a quantum state  $\ket{a_k}$ in the first part of the last register. Let the second part of the last register as the counter for each $a_k$. Then compare each $\ket{x_{I_j}^{(m)}} (m\in [M])$ with each potential configuration $a_k$,  the corresponding counter increased by  $1$ if they are identical, otherwise do nothing. We have
\begin{eqnarray}
\ket{\mathcal{S}_j }\ket{x_{I_j}^{(m)}}^{\otimes_{m=1}^M} \left(\ket{a_k}\ket{\textstyle\sum_m^M \mathbb{I}_{\{x^{(m)}_{I_j}=a_k\}}}\right)^{\otimes_{k=1}^{2^l}}.
\end{eqnarray}
\item[4).]  After dividing values in each counter by $M$,  we obtain
\begin{eqnarray}
\ket{\mathcal{S}_j }\ket{x_{I_j}^{(m)}}^{\otimes_{m=1}^M} \left(\ket{a_k}\ket{\widehat{p}(x_{I_j}=a_k)}\right)^{\otimes_{k=1}^{2^l}}.
\end{eqnarray}
\item[5).]  Undo step 2, it then results in the state in Eq.~(\ref{eq_quantum_state_probability}).
\end{itemize}

We now analyze the time complexity.  By Definition \ref{MRF_input}, Step 2 requires $\Ord{Ml\log n}$ for $M$ samples of $l$ nodes.  Step 3 compares each of $M$ samples with each of $2^l$ configurations, and each configuration involving $l$ numbers, then the time complexity is $\Ord{lM2^l}$.  It takes $\Ord{2^l}$ time to perform the division operations. The last step requires the same time as Step 2.  Therefore, in total the run time should be $$\Ord{lM2^l+Ml\log n}\subset \tOrd{2^{r-1}M+M\log n}$$  as $l\leq r-1$. 
\end{proof}

We then turn to calculate $\widehat{p}(x_{I\cup S})$ and $\widehat{p}(x_{I\cup S\cup\{u\}})$   for each potential configurations. For set $S$, assume there are  $s$ different configurations in the $M$ samples  which we denote as $\{b_1,b_2,\cdots,b_s\}$.  Notice that  $s\leq \min\{2^{\vert S \vert }, M \}$. 
 There are thus $s2^l$ and $s2^{l+1}$ potential different configurations of subset $\left(S\cup I\right)$ and $\left(I\cup S\cup \{u\}\right)$. Because the set $S$ and node $u$ are fixed, the method to calculate  $\widehat{p}(x_{I\cup S})$ and $\widehat{p}(x_{I\cup S\cup\{u\}})$ are very similar. We give the method and run time to prepare the empirical probability $\widehat{p}(x_{I\cup S})$ in the following. 

\begin{lemma}\label{lemma_empirical_probability_two_sets}
Suppose we are given quantum access to  $M$ samples of an MRF as Definition \ref{MRF_input}.  Fix a node $u\in [n]$, and a set $S\subset [n]\setminus \{u\}$ with $s$ configurations in the $M$ samples,   a string $\mathcal S_j=\left(j_1,j_2,\cdots j_l\right)\in \mathcal H_l$ corresponding to a subset $I_j \subseteq [n] \setminus\{u \cup S\}$ with size $l \in [r-1]$, for each configuration  $x_{(I_j\cup S) } =\left( a_i,b_t\right)$ where $a_k \in \{-1,1\}^l$  for $k\in [2^l]$ and  $b_t \in \{-1,1\}^{\vert S \vert}$  for $t\in [s]$.  There is a unitary $U_{I\cup S}$ operator which performs
\begin{eqnarray}
 & &\ket{\mathcal{S}_j }  \ket{\bar{0}}^{\otimes {(2^l+s)} }  \ket{\bar{0}}^{\otimes {s2^l} } \nonumber\\
  &\to &\ket{\mathcal{S}_j } \ket{\textbf{a}} \ket{\textbf{b}}\left(\otimes_{k=1}^{2^l} \otimes_{t=1}^{s} \ket{\widehat{p}\left( a_k, b_t\right)}\right)  \label{eq_quantum_state_probability_two_sets}
\end{eqnarray}
in time $\tOrd{ \vert S \vert M\left(\log n  + s2^{r-1}\right) }$, where $\ket{\textbf{a}}=\ket{a_k}^{\otimes_{k=1}^{2^l} } $, $\ket{\textbf{b}}=\ket{ b_t}^{\otimes_{t=1}^s} ,$  and $\ket{\widehat{p}\left(a_k, b_t\right)}$ is short for $  \ket{\widehat{p}\left(x_{I_j}=a_k, x_S=b_t\right)}.$ 
\end{lemma}

\begin{proof}
Recall that there are  $s$ different configurations for set $S$ and $2^l$ different configurations of subset $I$,  and thus $s2^l$  potential different configurations of subset $S\cup I$.   Performing unitary $U_{S\cup I}$ requires to compare the configuration of $S\cup I$ of each sample with the $s2^l$ potential configurations and obtaining the empirical probability $\widehat{p}\left(x_I,x_s\right)$ by a similar way of step 3 and 4 in the proof of Lemma \ref{lemma_probability_preparation}. The details of the procedure of preparing the state in  Eq.~(\ref{eq_quantum_state_probability_two_sets})  are shown in the following.
\begin{enumerate}[1).]
\item Prepare an initial state 
\begin{eqnarray}
\ket{\mathcal{S}_j } \ket{\bar{0}}^{\otimes M}\ket{\bar{0}} ^{\otimes {(2^l+s)}}  \ket{\bar{0}}^{\otimes {s2^l} },
\end{eqnarray}
where $\ket{\bar{0}}^{\otimes M}$ is used to store the $M$ samples of subset $I \cup S$,  $ \ket{\bar{0}}^{\otimes {(2^l+s)}} $ is used to store the $2^l$ different configurations of $I$ and $s$ different configurations of $S$, and $\ket{\bar{0}}^{\otimes {s2^l} }$ is used to store the empirical probability of each of the $2^ls$ configurations of subset $S\cup I.$
\item Query access to the $M$ samples as Definition \ref{MRF_input}  on the first two registers,  we obtain
\begin{eqnarray}
\ket{\mathcal{S}_j } \ket{\textbf{X}_{I_j\cup S }} \ket{\bar{0}}^{\otimes {(2^l+s)}} \ket{\bar{0}}^{\otimes {s2^l} },
\end{eqnarray}
where $ \ket{\textbf{X}_{I_j\cup S }} =\ket{X_{I_j\cup S }^{(m)}}^{\otimes_{m=1}^M}. $
\item Store the $2^l$ potential configures of $I_j$ and $s$ configurations of $S$ in the third register. Then compare each ${X_{I_j\cup S }^{(m)}}$ in the second register with each $\ket{a_k}\ket{b_t}$ in the third register, and store the sum of the indicator function of each configurations as step $3$ in the proof of Lemma \ref{lemma_probability_preparation},  we have
\begin{eqnarray}
\ket{\mathcal{S}_j } \ket{\textbf{X}_{I_j\cup S } }\ket{\textbf{a}}  \ket{\textbf{b}}\left( {\otimes_{k=1}^{2^l}}\otimes_{t=1}^s  \ket{ \Upsilon_{k,t}} \right),\nonumber
\end{eqnarray}
where $\ket{ \Upsilon_{k,t}}=\ket{\sum_m^M \mathbb{I}_{ \{x^{(m)}_{I_j}=a_k, x^{(m)}_S=b_t\}} }.$
\item  Divide each  $\Upsilon_{k,t}$  by $M$ and undo Step 2,  we have 
\begin{eqnarray}
\ket{\mathcal{S}_j }\ket{\textbf{a}}  \ket{\textbf{b}}\left( {\otimes_{k=1}^{2^l}}\otimes_{t=1}^s \ket{\widehat{p}\left(x_{I_j}=a_k, x_S=b_t\right)}\right).\nonumber
\end{eqnarray}
\end{enumerate}

We now turn to analyze the time complexity. Step $2$ requires time $\Ord{(l+\vert S \vert )M\log n }$ as each query for one node takes time $\Ord{M\log n}$  by Definition \ref{MRF_input}, and there are $l+\vert S \vert $ nodes in set $I \cup S$.  In Step $3$,  it takes  $\Ord{s2^lM}$ comparisons of each configuration with the $M$ samples, and each comparison involving $l+\vert S \vert$ numbers. Thus Step $3$ takes time $\Ord{(l+\vert S \vert) s2^lM}$. The division is on all  $\Upsilon_{k,t}$ for $k\in [2^l], t\in [s]$  in step 4, the run time of this step is then $\Ord{ s2^l + (l+\vert S \vert )M\log n}$ as it contains an undo operation of Step $2$. Therefore, the total run time  for preparing state in Eq.~(\ref{eq_quantum_state_probability_two_sets}) is 
\begin{eqnarray}
& &\Ord{ \left(l+\vert S \vert \right)M\log n +\left(l+\vert S \vert\right) s2^lM }\nonumber\\
&=&\tOrd{ \vert S \vert M\left(\log n  + s2^{r-1}\right) },
\end{eqnarray}
as $l\in[r-1].$
\end{proof}

Recall that the quantity $\hat{v}_{u, {I_j} \vert S}$ is related to empirical probabilities $\widehat{p}\left(X_u=1\right)$, $\widehat{p}\left(X_u=-1\right)$, $\widehat{p}(x_u,x_I,x_s)$, $\widehat{p}\left(x_u,x_s\right)$, $\widehat{p}\left(x_I,x_s\right)$, $\widehat{p}(x_s)$, and $\widehat{p}(x_I)$ of each configuration as shown in Eq.~(\ref{eq_expand_version_v}).  Therefore, via preparing these probabilities in quantum states by Lemma \ref{lemma_probability_preparation}  and  Lemma \ref{lemma_empirical_probability_two_sets}, after summing over all configurations accordingly,  a quantum state representing $\hat{v}_{u, {I_j} \vert S}$ in the superposition of all potential subset $I$ with size $l$ can be obtained. The method and required run time are given in the following lemma. 

\begin{lemma}\label{lemma_quantum_information_v}
Given quantum access to $M$ samples of an MRF as Definition \ref{MRF_input}, fix $l\in[r-1]$ and a node $u\in [n]$,  subset $S\subset [n]\setminus \{u\} $ with size $\vert S \vert <L$, and string $\mathcal S_j=\left(j_1,j_2\cdots j_l\right), j_k \in [n] \setminus\left(\{u\}\cup S\right)$ for	$k\in [l]$, there is a unitary operator $U_{v}$ which performs 
\begin{eqnarray}
\ket{0}\ket{\bar{0}}\ket{\bar{0}} \to 
\frac{1}{\sqrt{2K'}} \left(\textstyle\sum_{\mathcal S_j\in \mathcal H_l} \ket{1} \ket{\mathcal S_j}\ket{\hat{v}_{u, {I_j} \vert S}}  + \ket{\phi} \right)\nonumber\\ \label{eq_quantum_information_v}
\end{eqnarray} 
in time $ \tOrd{\vert S \vert M\left(s{2^{r-1}}+\log n\right)}$, where  the normalization factor $K'= (2^{\lceil\log n \rceil})^{r-1}$, $I_j=I(\mathcal S_j)$ is the subset mapping from string $\mathcal S_j,$ and $ \ket{\phi}= \sum_{\mathcal S_j\notin \mathcal H_l} \ket{0} \ket{\mathcal S_j} \ket{\bar{0}} $ .
\end{lemma}
\begin{proof}

We show the method to prepare state in Eq.~(\ref{eq_quantum_information_v}) first, then analyze the required run time. The procedure is shown in the following.     For simplicity, we ignore the  normalization factor $\frac{1}{\sqrt{2 K'}}.$
\begin{enumerate}[1).]
\item  Prepare an initial state  
$\ket{0}\left( \ket{0}^{\otimes\lceil \log n \rceil}\right)^{\otimes l},$
 then apply a Hadmard gate on each qubit except the first one,  we have 
\begin{eqnarray}
\textstyle  \sum_j\ket{0} \ket{\mathcal S_j}.
\end{eqnarray}
\item  Apply the unitary operator $U_{\mathcal H_l}$ on the state as  Lemma \ref{lemma_string_set_H_l}. We have 
\begin{eqnarray}
\textstyle \sum_{\mathcal{S}_j \in \mathcal H_l }\ket{1} \ket{\mathcal{S}_j }+ \sum_{\mathcal{S}_j \notin \mathcal H_l } \ket{0}\ket{\mathcal{S}_j }.
\end{eqnarray}
\item  Add enough quantum registers in state $\ket{\bar{0}}$ and apply controlled unitary operations $U_I$ as Lemma \ref{lemma_probability_preparation}, which controlled by the first register in state $\ket{1}$. We have
\begin{eqnarray}
\textstyle \sum_{\mathcal{S}_j \in \mathcal H_l }\ket{1} \ket{\mathcal{S}_j } \left( \ket{a_k} \ket{\widehat{p}\left(x_{I_j}=a_k\right)}\right) ^{\otimes_{k=1}^{2^l} } 
+ \ket{\Phi}.
\end{eqnarray}
where $\ket{\Phi}= \textstyle\sum_{\mathcal{S}_j \notin \mathcal H_l } \ket{0}\ket{\mathcal{S}_j }\ket{\bar{0}}.$
\item  Recall that the empirical probabilities $\widehat{p}\left(X_u=1\right)$, $\widehat{p}\left(X_u=-1\right)$, $\widehat{p}(x_s)$ and $\widehat{p}\left(x_u,x_s\right)$ can be obtained classically.  
The terms  $\widehat{p}(x_u,x_I,x_s)$, $\widehat{p}\left(x_I,x_s\right)$ can be obtained by performing controlled unitary operations $U_{I\cup S\cup \{u\}}$ and $U_{S \cup I }$ by Lemma \ref{lemma_empirical_probability_two_sets} which controlled by the first qubit in state $\ket{1}$. 
\item  After performing basic calculations based on Eq.~(\ref{eq_expand_version_v}), and undoing the controlled unitary operations, we obtain the state in Eq.~(\ref{eq_quantum_information_v}). 
\end{enumerate}

We now analyze the time complexity.  Step 1 costs $\Ord{ l\log n}$ time to apply the Hadamard gates.  According to Lemma \ref{lemma_string_set_H_l},  Step 2 takes $\Ord{l\log n}$ time.  For step 3, the bottleneck is to prepare state $\ket{\widehat{p}(x_u,x_I,x_s)}$ for each configuration of set $\{u\}\cup I \cup S$.  By Lemma \ref{lemma_empirical_probability_two_sets}, it takes time $\tOrd{\vert S \vert M\left(s{2^{r-1}}+\log n\right)}$. Step 5 costs $\Ord{ s2^{l}}$ time for the calculation and $\tOrd{\vert S \vert M\left(s{2^{r-1}}+\log n\right)}$ for the undoing operations.  Hence the total required run time should be
$ \tOrd{\vert S \vert M\left(s{2^{r-1}}+\log n\right)}.$ 
\end{proof}

\subsection{Subset finding by quantum maximum finding algorithm}
 After preparing the state in Eq.~(\ref{eq_quantum_information_v}),  a subset $I$ with size $l$ which satisfies $\hat{v}_{u, {I_j} \vert S}> \tau$  can be found by using quantum maximum finding algorithm \cite{durr1996quantum}, if there exists at least one such subset. The run time is given in the following Lemma. 
\begin{lemma}\label{lemma_quantum_search_subset}
Let $w>0, l\in[r-1]$. Suppose we are given quantum access to $M$ samples of an MRF with the same settings in Lemma  \ref{lemma_quantum_information_v} and the unitary operator $U_v$. 
\begin{itemize}
\item[•] If there exists subsets $I\in F_{u,S}^l$ satisfying  $\hat{v}_{u,I \vert S}>\tau$,  one such subset $I$  can be found in time $\tOrd{\vert S \vert s2^{r-1}M\sqrt{n^{r-1}}\log (1/w )}$ with success probability $1-w.$
\item[•] If there is no any subset $I\in F_{u,S}^l$  satisfying  $\hat{v}_{u,I \vert S}>\tau$, we can determine it in time $\tOrd{\vert S \vert s2^{r-1}M\sqrt{n^{r-1}}\log (1/w )}$  with success probability $1-w.$
\end{itemize}
\end{lemma}
\begin{proof}
Firstly, we find the maximum value of  $\hat{v}_{u,I \vert S}$ by using the quantum maximum finding algorithm \cite{durr1996quantum}.  If $\hat{v}_{u,I \vert S} > \tau$, we find a desired subset $I$. If the value is less than $\tau$, we know that there is no such subset. 

We now turn to analyze the run time.  According to Lemma \ref{lemma_quantum_information_v},  the run time for applying $U_v$  is $$C_1:=\tOrd{\vert S \vert M\left( s2^{r-1}+\log n\right)}.$$ As the quantum state in Eq.~(\ref{eq_quantum_information_v}) is a supposition of $2K'=\Ord{n^{r-1}}$ strings,   quantum maximum algorithm requires  $$C_2:=\tOrd{\sqrt{n^{r-1}}\log (1/w )}$$  calling of $U_v$  by Lemma \ref{quantum_max_finding}. Then it costs 
\begin{eqnarray}
C_1C_2 & = & \tOrd{ \vert S \vert M\left( s2^{r-1}+\log n\right) \sqrt{n^{r-1}}\log (1/w )  }\nonumber\\
&=&\tOrd{\vert S \vert s2^{r-1}M\sqrt{n^{r-1}}\log (1/w )}
\end{eqnarray}
 run time to find a subset $I$ with size $l\in [r-1]$ such that  $\hat{v}_{u, {I_j} \vert S}> \tau$.  
\end{proof}

\subsection{Main results }

 The quantum version of the classical algorithm \ref{MRF_information} is given as Algorithm \ref{algorithm_quantum_MRF_by_information}. Like the classical version, for each subset $S$, we start from finding a subset $I$ with size $r-1$ which satisfies $\hat{v}_{u, {I_j} \vert S}> \tau$. If there exists at least one such subset, we increase $S$ with subset $I$. Otherwise, if there is no such subset, we then search subsets with sizes $r-2$. In this way, we can find all these subsets. Until reach the bound $L$ of $S$,  we check each node $i$ in $S$,  remove node $i$ from $S$ if $\hat{v}_{u,i \vert S\setminus\{ i \}}< \tau$. The result is then the neighbours of node $u$.  The underlying graph therefore can be recovered via running the algorithm for every node.  
The time complexity and the number of samples required are shown in the following theorem. 

\begin{theorem}{ }
Fix $w > 0$. Suppose we are given quantum access to $M$ samples denoted as $X^{(m)}\in \{-1,1\}^n (m\in [M])$ from an $(\alpha,\beta)$-non-degenerate Markov random field with $r$-order interactions where the underlying graph has maximum degree at most $d$. Suppose that
 $$M \in \tOrd{\frac{2^{2L}}{\tau^2\delta^{2L}}(\log(1/w)+\log n)},$$
then with probability at least $1-w$, run quantum Algorithm \ref{algorithm_quantum_MRF_by_information} for node $u\in [n]$ recovers the correct neighborhood of $u$, and thus recovers the underlying graph $G$ by running Algorithm \ref{algorithm_quantum_MRF_by_information} for each node. Furthermore, the time complexity is  $\tOrd{2^{L}M\sqrt{n^{r+1}}\log \frac{1}{w}}.$
 \end{theorem}

\begin{figure}[htbp]
\begin{algorithm}[H]
  \caption{ \textsc{QuanMrfNbdh} }\label{algorithm_quantum_MRF_by_information}
  \begin{algorithmic}[1]
  \Require{Error $\epsilon\in (0,1)$, probability $w \in(0,1)$, quantum access to training set $\{X^{(m)} \in \{-1,1\}^n\} (m\in M) $ as quantum input \ref{MRF_input}, $r,w>0$
     }
    \State Fix input vertex $u$. Set $S:=\emptyset$.
    \While{$|S| \leq L$}
    \State $l\leftarrow r-1$
    \If{ find a subset $I \subset  [n]\setminus S$ of size  $l$ by using quantum maximum finding algorithm  with success probability $1-\frac{w}{2L(r-1)}$  as Lemma \ref{quantum_max_finding} such that
       $\hat{v}_{u,I\vert S} > \tau,$}
    \State set $S:=S\cup I$.
    \ElsIf{there is no such subset and $l>1$}
    \State $l\leftarrow  l-1$, go to step 4
    \EndIf
    \EndWhile
    \State For each $i \in S$, if   $\hat{v}_{u,i \vert S\setminus i }< \tau$,   then remove $i$ from $S$.
     \Ensure{ Set $S$  as our estimate of the neighborhood of $u$.}
  \end{algorithmic}
\end{algorithm}
\end{figure}
\begin{proof}
The sample complexity is the same as the classical algorithm. 

 For the time complexity, in Algorithm \ref{algorithm_quantum_MRF_by_information}, the bottleneck is Step $4$, which we find a subset $I$ by quantum maximum finding algorithm such that  $\hat{v}_{u,I\vert S > \tau}$.  By Lemma \ref{lemma_quantum_search_subset}, it takes time $$T_s:=\tOrd{\vert S \vert s 2^{r-1}M\sqrt{n^{r-1}}\log (1/w )}.$$  

As there are $L$ loops for at most $r-1$ inner loop,  the total run time of Algorithm \ref{algorithm_quantum_MRF_by_information} is then 
\begin{eqnarray}
\tOrd{rLT_s} & = & \tOrd{rL\vert S \vert s 2^{r-1}M\sqrt{n^{r-1}}\log (1/w )}\nonumber\\
&=& \tOrd{rL^2 2^{L}M\sqrt{n^{r-1}}\log (1/w )}\nonumber\\
& = & \tOrd{2^{L}M\sqrt{n^{r-1}}\log (1/w )},
\end{eqnarray}
as $l\leq r-1$, $\log s \leq  \vert S \vert \leq L$, and $2^r$ is a factor of $L$. 
Run Algorithm \ref{algorithm_quantum_MRF_by_information} for every node,  therefore it requires 
 \begin{eqnarray}
 \tOrd{2^{L}M\sqrt{n^{r+1}}\log (1/w ) }
 \end{eqnarray}
time to recover the structure of the underlying graph.  

We now analyze the success probability. The success probability for every quantum maximum finding is at least $p_s:=1-\frac{w}{2L(r-1)}$, then for $L$ loop, and each loop contains at most $r-1$ applications of the quantum maximum finding algorithm,  then by union bound, the total probability is  at least 
\begin{eqnarray}
 1-(1-p_s){L(r-1)}=1- \frac{w}{2}.
\end{eqnarray}
 Combining with the success probability of the original classical algorithm which we set as $1-\frac{w}{2}$, this leads to a total success probability  at least $1-w$ by using the union bound.  
\end{proof}
After obtaining the structure of the underlying graph, it is easy to learn the parameters of the MRFs with degree $d$ bounded underlying graph. \\

\section{Discuss and conclusion}\label{section_conclusion}

In this work, we proposed an efficient quantum algorithm for structure learning of $r$-wise MRFs with degree $d$ underlying graph,  based on the classical greedy algorithm \textsc{MrfNbhd} in Ref.~\cite{hamilton2017information} with a fixed search order. We also proved that the quantum algorithm is polynomial faster than the classical counterpart, in terms of the number of the variables, and the sample complexity remains the same.  More precisely, the quantum algorithm outperforms the classical algorithm if it satisfies 
\begin{eqnarray}
2^L M n^{\frac{r+1}{2}} \log \frac{1}{w} <  C_3Mn^r,
\end{eqnarray}
where $C_3>0$ is a constant, it yields
\begin{eqnarray}
n >C_4 \sqrt[r-1]{2^{2L}\log^2({1}/{w})},
\end{eqnarray}
for a constant $C_4>0$.

We believe the quantum structure learning algorithm introduced in this work has wide applications and inspires the construction of more efficient quantum structure learning algorithms.

\section{Acknowledgement}
This work is supported by the National Natural Science Foundation of China (Nos.62172341,61772437), and Shenzhen Institute for Quantum Science and Engineering, Southern University of Science and Technology (Grant No. SIQSE202105).

\bibliographystyle{apsrev}
\bibliography{MRF.bib}

\end{document}